\documentclass[11pt]{article}
\usepackage[utf8]{inputenc}
\usepackage{listings}
\usepackage{color}
\usepackage{longtable}
\usepackage{float}
\usepackage{amsmath}
\usepackage{authblk}
\newcommand{\pGF}{p(r, t \vert r_{0})}
\newcommand{\qGF}{q(r, t \vert r_{0})}
\newcommand{\sS}{\tilde{S}(s \vert r_{0})}

\newcommand{\tSbB}{S^{>}(t \vert \mathbf{B}_{2})}
\newcommand{\tSsB}{S^{<}(t \vert \mathbf{B}^{C}_{2})}

\newcommand{\kI}{k_{\text{irr}}(t)}

\title{}
\date{\today}

\begin{document}

\title{General theory of area reactivity models: rate coefficients, binding probabilities and all that}
\author{Thorsten Pr\"ustel} 
\author{Martin Meier-Schellersheim} 
\affil{Laboratory of Systems Biology\\National Institute of Allergy and Infectious Diseases\\National Institutes of Health}
\maketitle
\let\oldthefootnote\thefootnote 
\renewcommand{\thefootnote}{\fnsymbol{footnote}} 
\footnotetext[1]{Email: prustelt@niaid.nih.gov, mms@niaid.nih.gov} 
\let\thefootnote\oldthefootnote 
\abstract{
We further develop the general theory of the area reactivity model that provides an alternative description of the diffusion-influenced reaction of an isolated receptor-ligand pair in terms
of a generalized Feynman-Kac equation. We analyze both the irreversible and reversible reaction and derive the equation of motion for the survival and separation probability.
Furthermore, we discuss the notion of a time-dependent rate coefficient within the alternative model and obtain a number of relations between the rate coefficient, the survival and separation probabilities and the reaction rate. Finally, we calculate asymptotic and approximate expressions for the (irreversible) rate coefficient, the binding probability, the average lifetime of the bound state and discuss on- and off-rates in this context. Throughout our treatment, we will point out similarities and differences between the area and the classical contact reactivity model. The presented analysis and obtained results provide a theoretical framework that will facilitate the comparison of experiment and model predictions.     
}
\section{Introduction}
\label{sec-1}

Recently, the volume and area reactivity (AR) model in three \cite{Erban:2011, Khokhlova:2012BullKorCS, Khokhlova:2012} and two dimensions (2D) \cite{Prustel_Area:2014}, respectively, have been proposed as an alternative framework for the description of the diffusion-influenced reaction of an isolated pair. Typically, microscopic models that provide a more detailed description than a rate equation approach, depict a diffusion-influenced reactions as two-step processes where the molecules' diffusive motion is a prerequisite for the actual reaction and hence becomes an essential element of the theoretical treatment \cite{Goesele:1984, Rice:1985}. The classical Smoluchowski approach asserts that when the molecules approach each other at a critical distance $r=a$ (referred to as \textit{encounter} or \textit{reaction} radius), the actual reaction may occur with a certain probability. Therefore, we will henceforth refer to Smoluchowski-type models as contact reactivity (CR) models. This picture is implemented in the mathematical description by imposing certain boundary conditions (BC) on the solutions of the Smoluchowski equation that incorporate the physics at the encounter distance. The irreversible reaction is implemented by the radiation boundary condition that involves an intrinsic association constant $\kappa_{a}$ \cite{collins1949diffusion}. The radiation BC generalizes the classical Smoluchowski (\textit{absorbing}) BC \cite{smoluchowski:1917}, that corresponds to the limit $\kappa_{a} = \infty$, meaning that every encounter leads to a reaction. Reversible reactions can be incorporated through another generalization known as \textit{backreaction} BC \cite{Goodrich:1954, Agmon:1984, kimShin:1999, TPMMS_2012JCP} that takes into account dissociations and includes an additional intrinsic dissociation constant $\kappa_{d}$. 

In contrast to CR models, the AR model abandons the assumption of the existence of a sharply defined reaction radius. Instead, it is based on the idea that the reaction can occur throughout 
a reaction area. Mathematically, this model is implemented by a generalized version of the Feynman-Kac equation (FKE) \cite{Khokhlova:2012BullKorCS, Khokhlova:2012,Prustel_Area:2014}.
Within this model, BC play no role to incorporate the actual reaction, instead sink terms \cite{Wilemski:1973} are added to the diffusion equation to accomodate the interactions. Exact expressions for the Green's function, survival and binding probabilities in the Laplace (3D case) \cite{Khokhlova:2012BullKorCS, Khokhlova:2012} and in both the Laplace and time domain (in 2D) have been derived \cite{Prustel_Area:2014}. However, compared to the degree of maturity of CR theories \cite{Goesele:1984, Rice:1985, Agmon:1990p10}, the AR model still lacks a fully developed framework and a more unified treatment is missing.
   
The manuscript's goal is to address this need. In the next section, we will focus on the irreversible reaction and introduce our notation. Starting from the underling FKE, we will     
derive an equation of motion for the survival probabiliy. Next, we will deal with the time-dependent rate coefficent that plays a central role in the Smoluchowski theory and will derive a number of relations that resemble the situation in CR theories. Then, we calculate approximate expressions for the rate coefficient for short and large times.
In the following sections, we switch to the reversible case and proceed analogously to the irreversible case. In addition, we will discuss the average lifetime of the bound state that gives rise to the off-rate and we will obtain approximate expressions for it.
\section{Irreversible reaction}
\label{sec-2}

We consider a pair of molecules $A$ and $B$ that undergo a diffusive motion characterized by diffusion constants $D_{A}$ and $D_{B}$, respectively. Equivalently, one may view this system
as a single molecule diffusing with diffusion constant $D=D_{A}+D_{B}$ around a static molecule that, without loss of generality, is assumed to be located at the origin. According to the AR model, the molecule may bind whenever its position is located within the static reaction disk of radius $r=a$. The probability density function (PDF) $\pGF$ gives the likelihood of finding
the molecule unbound at a distance $r$ at time $t$, provided that its distance was initially $r_{0}$ at time $t=0$. There are two main differences to the CR model. First, the diffusing molecule may be located within the reaction area without being bound, hence, the PDF $\pGF$ is also defined for $r<a$. Second, the bound state is infinitely degenerate. Therefore, it is natural to introduce another PDF $q(r,t\vert r_{0})$ that gives the probability to find the molecule bound at a distance $r$ at time $t$, given that it was unbound at a distance $r_{0}$ at time zero. The equations of motion of $\pGF$ and $\qGF$ are \cite{Khokhlova:2012}
  \begin{eqnarray}
  \frac{\partial p(r,t\vert r_{0})}{\partial t} &=& \mathcal{L}_{r}p(r,t\vert r_{0}) - \kappa_{r}\Theta(a - r)p(r,t\vert r_{0}), \label{irr_FKE_p} \\
  \frac{\partial q(r,t\vert r_{0})}{\partial t} &=& \kappa_{r}\Theta(a - r)p(r,t\vert r_{0}), \label{irr_FKE_q}
  \end{eqnarray}
where $\Theta(x)$ denotes the Heaviside step-function that assumes unity for $x>0$ and vanishes otherwise. We note that the term $\kappa_{r}\Theta(a - r)p(r,t\vert r_{0})$
describes the association reaction and $\mathcal{L}_{r}$ denotes the 2D rotationally symmetric diffusion operator
\begin{equation}\label{def_diffusion_operator}
\mathcal{L}_{r} = D\frac{1}{r}\frac{\partial}{r}\bigg(r\frac{\partial}{r}\bigg).
\end{equation}
Eq.~(\ref{irr_FKE_p}) is referred to as the Feynman-Kac equation.
We see that, in the irreversible case, the equations of motion for $\pGF$ and $\qGF$ [Eqs.~(\ref{irr_FKE_p}),~(\ref{irr_FKE_q})] are decoupled and that the equation for the PDF $\qGF$ is an ordinary differential equation \cite{Khokhlova:2012}. In fact, knowledge of $\pGF$ allows to obtain $\qGF$ easily via Eq.~(\ref{irr_FKE_q}).
The initial conditions (IC) are
  \begin{eqnarray}
  p(r,t=0 \vert r_{0}) &=& \frac{\delta(r - r_{0})}{2\pi r_{0}}, \label{IC_FKE_p}\\
  q(r,t=0 \vert r_{0}) &=& 0.  \label{IC_FKE_q}
  \end{eqnarray}
The FKE Eq.~(\ref{irr_FKE_p}) is subject to BC at the origin and at infinity, respectively:
  \begin{eqnarray}
    r\frac{\partial}{\partial r} p(r,t\vert r_{0}) && \underset{r \rightarrow 0}{\longrightarrow} 0,  \label{BC_origin} \\
    p(r,t\vert r_{0}) && \underset{r \rightarrow \infty}{\longrightarrow} 0. \label{BC_infinity}
  \end{eqnarray}

An important quantity that can be derived from the GF is the survival probability $S(t\vert r_{0})$
  \begin{equation}\label{def_survival_probability}
    S(t \vert r_{0}) = 2\pi \int^{\infty}_{0} p(r,t \vert r_{0}) r dr
  \end{equation}
that gives the likelihood that a pair, initially separated by $r_{0}$ does not bind by time $t$.
From the IC Eq.~(\ref{IC_FKE_p}) it follows that
  \begin{equation}
    S(t = 0 \vert r_{0}) = 1.
  \end{equation}
It will turn out to be useful to introduce additional “survival probabilities” $S^{<}(t \vert r_{0}), S^{>}(t \vert r_{0})$
  \begin{eqnarray}
    S^{<}(t \vert r_{0}) = 2\pi \int^{a}_{0} p(r,t \vert r_{0}) r dr, \label{eq_def_S<_irr}\\
    S^{>}(t\vert r_{0}) = 2\pi \int^{\infty}_{a} p(r,t \vert r_{0}) r dr, \label{eq_def_S>_irr}
  \end{eqnarray}
that give the joint probability that by time $t$ the molecule has not reacted yet and that it is located within/outside the reaction area, respectively, given that it was initially located at a distance $r_{0}$.   
Obviously, one has
  \begin{equation}\label{def_S_sum}
  S(t\vert r_{0}) = S^{<}(t\vert r_{0}) + S^{>}(t\vert r_{0}).
  \end{equation}
The corresponding IC read
  \begin{eqnarray}
    S^{<}(t = 0 \vert r_{0}) &=& \Theta(a-r_{0}), \label{IC_S_<}\\
    S^{>}(t = 0 \vert r_{0}) &=& \Theta(r_{0} - a). \label{IC_S_>}
  \end{eqnarray}

Now, by integrating the FKE Eq.~(\ref{irr_FKE_p}) over all $r$ from the origin to infinity, multiplying by $2\pi$ and using the BC Eqs.~(\ref{BC_origin}),~(\ref{BC_infinity}), we obtain
  \begin{equation}\label{eq_S}
    \frac{\partial S(t \vert r_{0})}{\partial t}  = -\kappa_{r} S^{<}(t \vert r_{0}),
  \end{equation}
Note that one may view Eq.~(\ref{eq_S}) as an analogue of the relation
  \begin{equation}
    \frac{\partial S(t \vert r_{0})}{\partial t} = -\kappa_{a} p(a, t\vert r_{0}),
  \end{equation}
known from CR theories \cite{Agmon:1990p10}.

The binding probability is defined by
  \begin{equation}\label{def_binding_probability}
    Q(t\vert r_{0}) = 2\pi\int^{\infty}_{0}\qGF r dr = 2\pi\int^{a}_{0}\qGF r dr, 
  \end{equation}
where the last equation does hold because $\qGF = 0$ for $r > a$. For consistency, one requires $S(t \vert r_{0}) + Q(t \vert r_{0}) = 1$ for all times. In fact, this follows from
the equations of motion Eqs.~(\ref{irr_FKE_p}),~(\ref{irr_FKE_q}), the IC (Eqs.~(\ref{IC_FKE_p}),~(\ref{IC_FKE_q})) and the definitions of the survival and binding probabilities (Eqs.~(\ref{def_survival_probability}),~(\ref{def_binding_probability})), cf. Ref.~\cite{Prustel_Area:2014}.

The reaction rate gives the fraction of molecule pairs that associate with each other per unit time and is defined by the negative time derivative of the survival probability \cite{Agmon:1990p10}
\begin{equation}\label{def_R}
R(t\vert r_{0}) = -\frac{\partial S(t\vert r_{0})}{\partial t},
\end{equation}
which translates to
  \begin{equation}\label{eq_def_R_laplace}
    \tilde{R}(s \vert r_{0}) = 1 - s\tilde{S}(s \vert r_{0})
  \end{equation}
in the Laplace domain. In general, we employ the following notation for the  Laplace transform of a function $f(t)$:
  \begin{equation}
    \mathbf{L}[f(t)](s) = \tilde{f}(s) = \int^{\infty}_{0}e^{-st}f(t)dt. 
  \end{equation}
It follows from Eq.~(\ref{eq_S}) and (\ref{def_R}) that
  \begin{equation}\label{R_and_area}
    R(t \vert r_{0}) = \kappa_r S^{<}(t \vert r_{0}).
  \end{equation}
In the context of the classical Smoluchowski theory one has the analogous relation \cite{Agmon:1990p10}
  \begin{equation}
    R(t \vert r_{0}) = -J(a, t \vert r_{0}) = \kappa_{a} p(a, t\vert r_{0}),
  \end{equation}
that shows that reactions only take place at contact $r=a$. The central postulate of the AR model that the reaction may occur throughout an reaction area is reflected in Eq.~(\ref{R_and_area}).

To obtain an equation of motion for the survival probability, we start from the FKE Eq.~(\ref{irr_FKE_p}), make use of the detailed balance condition
\begin{equation}\label{detailed_balance}
\pGF = p(r_{0}, t \vert r)
\end{equation}
and switch $r\leftrightarrow r_{0}$, which results in
    \begin{equation}\label{backward_GF}
  \frac{\partial p(r, t\vert r_{0})}{\partial t} = \mathcal{L}_{r_{0}}p(r, t\vert r_{0}) - \kappa_{r}\Theta(a - r_{0})p(r, t\vert r_{0}),
    \end{equation}
Finally, we integrate Eq.~(\ref{backward_GF}) over all space $2\pi\int^{\infty}_{0}rdr$ to arrive at
  \begin{equation}\label{eq_motion_S}
    \frac{\partial S(t \vert r_{0})}{\partial t} = \mathcal{L}_{r_{0}}S(t \vert r_{0}) - \kappa_{r}\Theta(a - r_{0})S(t \vert r_{0}).
  \end{equation}
In the same way, again starting from Eq.~(\ref{backward_GF}), one can derive that both $S^{<}(t\vert r_{0})$ and $S^{>}(t\vert r_{0})$ obey an equation of motion that takes the same form as Eq.~(\ref{eq_motion_S}), but they satisfy different IC (Eqs.~(\ref{IC_S_<}),~(\ref{IC_S_>})). In this sense, the time evolution of $S^{<}(t\vert r_{0})$ and $S^{>}(t\vert r_{0})$ is decoupled.  

Now, the Laplace transform of Eq.~(\ref{eq_motion_S}) reads \cite{Khokhlova:2012}
  \begin{equation}
    \mathcal{L}_{r_{0}}\tilde{S}(s \vert r_{0}) = [\kappa_{r}\Theta(a-r_{0})+s]\sS - 1,
  \end{equation}
which may be rewritten more explicitly as,
\begin{equation}\label{laplace_operator_S}
     \mathcal{L}_{r_{0}}\tilde{S}(s \vert r_{0}) =
    \begin{cases}
      D w^{2} \sS - 1 & \text{if} \quad r_{0} < a, \\
      D v^{2} \sS - 1 & \text{if} \quad r_{0} > a,
    \end{cases}
\end{equation}
where we defined
  \begin{eqnarray}
    v &=& \sqrt{\frac{s}{D}}, \label{eq_def_v_irr} \\
    w &=& \sqrt{\frac{s+\kappa_{r}}{D}}, \label{eq_def_w_irr} 
  \end{eqnarray}
From Eq.~(\ref{laplace_operator_S}) and the Laplace transform  of the reaction rate (Eq.(\ref{eq_def_R_laplace})), we obtain
  \begin{equation}\label{laplace_operator_R}
      \mathcal{L}_{r_{0}} \tilde{R}(s \vert r_{0}) =
      \begin{cases}
        Dw^{2} \tilde{R}(s \vert r_{0}) - \kappa_{r} & \text{if} \quad r_{0} < a, \\
        Dv^{2} \tilde{R}(s \vert r_{0})  & \text{if} \quad r_{0} > a.
      \end{cases}
  \end{equation}
We again note that Eqs.~(\ref{laplace_operator_S}),~(\ref{laplace_operator_R}) have counterparts in the classical Smoluchowski theory \cite{Agmon:1990p10}.

Finally, we would like to derive relations between the different survival probabilities $S(t \vert r_{0}), S^{<}(t \vert r_{0}), S^{>}(t \vert r_{0})$.
To this end, we start from Eq.~(\ref{irr_FKE_p}) to obtain 
  \begin{eqnarray}
    \frac{\partial S^{>}(t \vert r_{0})}{\partial t} &=& -2\pi a D \frac{\partial \pGF}{\partial r}\vert_{r=a}, \label{eq_S>} \\
    \frac{\partial S^{<}(t \vert r_{0})}{\partial t} &=& 2\pi a D \frac{\partial \pGF}{\partial r}\vert_{r=a} -\kappa_{r} S^{<}(t \vert r_{0}). \label{eq_S<}
  \end{eqnarray}
Therefore, we can conclude that
  \begin{equation}
    S^{>}(t \vert r_{0}) = \Theta(r_{0} - a) - 2\pi a D \int^{t}_{0}\frac{\partial p(r, \tau \vert r_{0})}{\partial r}\vert_{r=a} d\tau,
  \end{equation}
and, using the Laplace transforms of Eqs.~(\ref{eq_S>}),~(\ref{eq_S<}), we arrive at 
  \begin{eqnarray}
    \tilde{S}^{<}(s \vert r_{0}) &=& \frac{1}{s+\kappa_{r}} - \frac{s}{s+\kappa_{r}}\tilde{S}^{>}(s \vert r_{0}), \\
    \tilde{S}(s \vert r_{0}) &=& \frac{1}{s+\kappa_{r}} + \frac{\kappa_{r}}{s+\kappa_{r}}\tilde{S}^{>}(s \vert r_{0}).
  \end{eqnarray}
\subsection{Time-dependent rate coefficient}
\label{sec-2-1}

We now turn to the the time-dependent rate coefficent $k_{\text{irr}}(t)$ whose calculation is a central goal of the Smoluchowski theory. According to the conventional wisdom, $k_{\text{irr}}(t)$ is defined in the following way \cite{Agmon:1990p10}. 
Let $p(r,t \vert \text{eq})$ denote the radial distribution function of the reactants at time $t$, given that the initial distribution takes the 
equilibrium form 
  \begin{equation}
    p(r,t=0 \vert \text{eq}) = 
    \begin{cases}
      1 & \text{for} \quad r > a \\
      0 & \text{for} \quad r < a.
    \end{cases}
  \end{equation}
Then, within the framework of CR theories, the time-dependent rate coefficient is defined as the reactive flux at the encounter distance
  \begin{equation}\label{def_k}
    k_{\text{irr}}(t) = 2\pi a D \frac{\partial p(r, t \vert \text{eq})}{\partial r}\vert_{r=a}
  \end{equation}
Already in the appendix of Ref.~\cite{Prustel_Area:2014} it was discussed that this definition might have to be reconsidered for theories that abandon the notion of an encounter radius and that instead assume that the reaction can occur throughout an interaction area. Hence, we are interested in alternative expressions for $k_{\text{irr}}(t)$ that may be viewed as more general than Eq.~(\ref{def_k}) in the sense that they are suitable for both CR and AR models.
An alternative way to calculate $k_{\text{irr}}(t)$ is \cite{Agmon:1990p10}
  \begin{equation}\label{kI_by_R}
    k_{\text{irr}}(t) = 2\pi \int^{\infty}_{a} R(t\vert r_{0}) r_{0} dr_{0}.
  \end{equation}
Now, we can proceed exactly as within the framework of CR theories: 
Because the reaction rate is the negative time derivative of the survival probability (Eq.~(\ref{def_R})),
one has
  \begin{eqnarray}
    k_{\text{irr}}(t) &=& -2\pi\int_{a}^{\infty} \frac{\partial S(t\vert r_{0})}{\partial t} r_{0} dr_{0}  \label{kI_by_dt_S} \\
&=& 2 \pi a D \frac{\partial S(t \vert r_{0})}{\partial r_{0}}\vert_{r_{0}=a}. \label{kI_by_dr_S}
  \end{eqnarray}
The second identity follows upon integrating the equation of motion of the survival probability (Eq. (\ref{eq_motion_S})) over $r_{0}$ from $a$ to $\infty$.
We emphasize that Eqs.~(\ref{kI_by_R}),~(\ref{kI_by_dt_S}) and (\ref{kI_by_dr_S}) do hold in exactly the same form, both in CR and AR theories.

In the classical case, due to the radiation BC, one can also relate $k_{\text{rad}}(t)$ with the survival probability at contact \cite{Agmon:1990p10}
  \begin{equation}\label{kI_cr_Sa}
    k_{\text{rad}}(t) = \kappa_{a} S(t\vert a).
  \end{equation}
There is an analogous relation to Eq.~(\ref{kI_cr_Sa}) in the context considered here. Using Eq.~(\ref{eq_S}) one arrives at
  \begin{eqnarray}
    \kI &=& \kappa_{r} \tSsB \\
        &=& \kappa_{r} \tSbB \label{k_irr_S},
  \end{eqnarray}
where we have introduced
  \begin{eqnarray}
    \tSsB &=& 2\pi \int^{\infty}_{a} S^{<}(t \vert r_{0}) r_{0} dr_{0} = 4\pi^{2} \int^{\infty}_{a} \int^{a}_{0} \pGF r \, dr \, r_{0} \, dr_{0},\quad\quad \label{def_SsB} \\
    \tSbB &=& 2\pi \int^{a}_{0} S^{>}(t \vert r_{0}) r_{0} dr_{0} = 4\pi^{2} \int^{a}_{0} \int^{\infty}_{a} \pGF r \, dr \, r_{0} \, dr_{0}. \quad\quad\label{def_SbB}
  \end{eqnarray}
Obviously, it follows that $\tSsB = \tSbB$. Note that $\tSsB$, $\tSbB$ are technically no probabilities, but it is natural to relate these quantities to the probability
  \begin{equation}
    S^{>}(t \vert \, \text{uni}) =   \frac{1}{\pi a^{2}} S^{>}(t \vert \mathbf{B}_{2}),
  \end{equation}
that the molecule is found unbound with $r>a$ at time $t$, given that initially it was uniformly
distributed inside the reaction area. Then, we can write
  \begin{eqnarray}
    \kI &=& \pi a^{2} \kappa_{r} S^{>}(t \vert \, \text{uni}).
  \end{eqnarray}
As in the CR case, one can relate $\kI$ and the time-dependent rate coefficient associated with absorbing BC. To this end, we follow the argumentation from Ref.~\cite{Agmon:1990p10}. 
The reaction rate $R_{\text{abs}}(t' \vert r_{0})$ corresponding to absorbing BC gives the rate of molecules that arrive at contact for the first time at $t' < t$. The AR model's time-dependent reaction rate at contact $R_{\text{irr}}(t - t' \vert r_{0}=a)$ yields the likelihood that the molecules react during $t - t'$. Therefore, one has the convolution relation 
  \begin{equation}\label{conv_R}
    R(t \vert r_{0}) = \int^{t}_{0} R(t-t' \vert a)R_{\text{abs}}(t' \vert r_{0}) dt'.  
  \end{equation}
Taking into account the definition of the reaction rate (Eq.~(\ref{def_R})) and Eq. (\ref{eq_motion_S}), the relation Eq.~(\ref{conv_R}) leads to
  \begin{equation}
    \frac{\partial \tilde{S}(s \vert r_{0})}{\partial r_{0}}\vert_{r_{0}=a} = \tilde{R}(s \vert a)\frac{\partial\tilde{S}_{\text{abs}}(s \vert r_{0})}{\partial r}\vert_{r_{0}=a},  
  \end{equation}
and hence one arrives at
  \begin{equation}
    \tilde{k}_{\text{irr}}(s) = \tilde{R}(s\vert a)\tilde{k}_{\text{abs}}(s). 
  \end{equation}
\subsection{Asymptotic and approximate expressions for the rate coefficient}
\label{sec-2-2}
\subsubsection{Short time expansion}
\label{sec-2-2-1}
To derive a short time asymptotic expansion of $k_{\text{irr}}(t)$, we start from the Laplace domain expression for the reversible time-dependent rate coefficient \cite{Prustel_Area:2014} and set $\kappa_{d} = 0$ to obtain
  \begin{equation}\label{kI_explicit}
    \tilde{k}_{\text{irr}}(s) = \frac{2 \pi a \kappa_{r}}{D}\frac{1}{vw}\frac{K_{1}(va)I_{1}(wa)}{\mathcal{N}},
  \end{equation}
where 
  \begin{equation}
    \mathcal{N} = v I_{0}(wa)K_{1}(va) + w I_{1}(wa)K_{0}(va).
  \end{equation}
Note that here $v,w$ are defined by Eqs.~(\ref{eq_def_v_irr}),~(\ref{eq_def_w_irr}).

Because $t \rightarrow 0$ in the time domain corresponds to $s \rightarrow \infty$ in the Laplace domain, we employ the large argument expansions of the modified Bessel functions to obtain \cite[Appendix III, Eqs. (11),(12)]{carslaw1986conduction}
  \begin{equation}
    \tilde{k}_{\text{irr}}(s) = \pi a \kappa_{r} \sqrt{D} \frac{1}{s^{3/2}}\bigg[1 - \bigg(\frac{3}{8}\frac{D}{a^{2}} + \frac{3}{4}\kappa_{r}\bigg)\frac{1}{s} \bigg].
  \end{equation}
To convert this expression to its time domain counterpart, we employ 
\cite[Appendix V, Eq.(2)]{carslaw1986conduction}
  \begin{equation}
    \mathbf{L}^{-1}\bigg(\frac{1}{s^{\nu+1}}\bigg) = \frac{t^{\nu}}{\Gamma(\nu+1)}.
  \end{equation}  
Using $\Gamma(3/2) = 1/2\sqrt{\pi}, \, \Gamma(5/2) = 3/4\sqrt{\pi}$, we arrive at
  \begin{equation}\label{eq_kirr_small_time}
    \kI = \sqrt{4\pi Dt} a\kappa_{r} \bigg[1 - \frac{1}{2}\bigg(\frac{1}{2}\frac{D}{a^{2}} + \kappa_{r}\bigg) t + \ldots \bigg].
  \end{equation}
It is instructive to compare this expression with the CR results for absorbing and radiation BC \cite{Barzykin:1993}, respectively
  \begin{eqnarray}
    k_{\text{abs}}(t) &=& 2\pi D \bigg[ \frac{1}{\sqrt{\pi \tau}} + \frac{1}{2} - \frac{1}{4}\sqrt{\frac{\tau}{\pi}} + \ldots\bigg], \label{eq_kabs_small_time} \\ 
    k_{\text{rad}}(t) &=& \kappa_{a} \bigg[ 1 - 2\tilde{h}\sqrt{\frac{\tau}{\pi}} + \tilde{h}\bigg(\tilde{h} + \frac{1}{2}\bigg)\tau + \ldots\bigg], \label{eq_krad_small_time}  
  \end{eqnarray}
where $\tau = D/a^{2} t$, $\tilde{h} = \kappa_{a}/(2\pi D)$.
We emphasize the following points. First, the AR model gives a rate coefficient that does not possess a singularity at $t=0$, similar to the CR model with radiation BC, but in contrast to the case of absorbing BC. Second, the AR model predicts that $k_{\text{irr}}(t)$ vanishes at $t=0$, whereas the CR model with radiation BC gives $k_{\text{rad}}(t=0) = \kappa_{a}$. Third, it follows from Eqs.~(\ref{eq_kirr_small_time}),~(\ref{eq_krad_small_time}) that
  \begin{eqnarray}
    \int^{t}_{0}k_{\text{irr}}(\tau)d\tau &\sim& t^{3/2}, \\
    \int^{t}_{0}k_{\text{rad}}(\tau)d\tau &\sim& t,
  \end{eqnarray}
which implies that for small times the time integral over the time-dependent rate coefficient grows slower in the AR case than in the CR case.
\subsubsection{Large time expansion}
\label{sec-2-2-2}

We are now interested in $\kI$ for long times $t\rightarrow\infty$. 
Again, using Eq. (\ref{kI_explicit}) as a starting point and the expansion of the modified Bessel functions \cite[Appendix III, Eqs.(7),(10)]{carslaw1986conduction}, we arrive at
  \begin{equation}
    \tilde{k}_{\text{irr}}(s) = -4\pi D \frac{1}{s\ln\bigg(\frac{1}{4D}a^{2}e^{2(\gamma-\rho)}s\bigg)},
  \end{equation}
where $\gamma = 0.57722\ldots$ denotes the Euler-Mascheroni constant \cite{abramowitz1964handbook}
and 
  \begin{equation}
    \rho = \frac{I_{0}(\sqrt{\kappa_{r}/D}a)}{I_{1}(\sqrt{\kappa_{r}/D}a)}\sqrt{\frac{D}{a^{2}\kappa_{r}}}.
  \end{equation}
For the inversion to the time domain we employ
  \begin{equation}
    -\mathbf{L}^{-1}\bigg[\frac{1}{s\ln{(Cs)}}\bigg] = \frac{1}{\ln(C^{-1} t)} - \frac{\gamma}{(\ln C^{-1}t)^{2}} + \ldots,
  \end{equation}
where $C>0$ denotes an arbitray constant
to obtain finally
  \begin{equation}
    \kI = 4\pi D \bigg(\frac{1}{\ln(4 \tau -2\gamma+2\rho)} - \frac{\gamma}{[\ln(4 \tau -2\gamma+2\rho)]^{2}} + \ldots\bigg).\label{eq_large_time_k_irr}
  \end{equation}
\section{Reversible reactions}
\label{sec-3}

For the reversible reaction, the equations of motion of the PDF $p_{\text{rev}}(r,t\vert r_{0})$ and $q_{\text{rev}}(r,t\vert r_{0})$ need to be extended to allow for dissociation of the bound pair \cite{Khokhlova:2012}
  \begin{eqnarray}
    \frac{\partial p_{\text{rev}}(r,t\vert r_{0}) }{\partial t} &=& \mathcal{L}_{r}p_{\text{rev}}(r,t\vert r_{0}) - \kappa_{r}\Theta(a - r)p_{\text{rev}}(r,t\vert r_{0}) + \kappa_{d}q_{\text{rev}}(r,t\vert r_{0}), \label{rev_FKE_1}  \quad\quad \\
    \frac{\partial q_{\text{rev}}(r,t\vert r_{0}) }{\partial t} &=& \kappa_{r}\Theta(a - r)p_{\text{rev}}(r,t\vert r_{0}) - \kappa_{d}q_{\text{rev}}(r,t\vert r_{0}). \label{rev_FKE_2}
  \end{eqnarray}
Notice that the equations of motion are now coupled, in contrast to their irreversible cousins, due to the appearance of the term $\kappa_{d}q_{\text{rev}}(r,t\vert r_{0})$, which yields the rate of dissociation \cite{Khokhlova:2012}. Obviously, Eqs.~(\ref{rev_FKE_1}),~(\ref{rev_FKE_2}) reduce to their irreversible counterparts Eqs.~(\ref{irr_FKE_p}),~(\ref{irr_FKE_q}) for $\kappa_{d} \rightarrow 0$.

One requires that $p_{\text{rev}}(r,t\vert r_{0})$ is subject to the same BC at the origin and at infinity as its irreversible analogue (Eq.~(\ref{BC_origin}),~(\ref{BC_infinity})).

Regarding the IC, we have to be aware of the fact that for the reversible reaction also the bound pair represents a possible initial state. Therefore, besides the IC that describe the initially unbound molecule (Eq.~(\ref{IC_FKE_p}),~(\ref{IC_FKE_q})), we have to include in our description the following set of IC \cite{Khokhlova:2012}
  \begin{eqnarray}
    p_{\text{rev}}(r, t \vert r_{0}, \ast) &=& 0, \label{IC_FKE_p_rev}\\
    q_{\text{rev}}(r, t \vert r_{0}, \ast) &=& \frac{\delta(r - r_{0})}{2\pi r_{0}}. \label{IC_FKE_q_rev}
  \end{eqnarray}
Here, we make use of the notation adopted in Ref.~\cite{Prustel_Area:2014}, where the initially bound state is indicated by the symbol $\ast$ and the corresponding quantities like the GF solutions etc.~are denoted by $p_{\text{rev}}(r,t\vert r_{0}, \ast)$, $q_{\text{rev}}(r,t\vert r_{0}, \ast)$ and so forth.
   
Survival as well as binding probabilities and the reaction rate are techically defined in the same manner as in the irreversible case (Eqs.~(\ref{def_survival_probability}), (\ref{def_binding_probability}), (\ref{def_R}), (\ref{eq_def_S<_irr}), (\ref{eq_def_S>_irr})). However, we would like to point out that in the context of the reversible reaction the term ``survival probability'' is conceptually somewhat misleading and that, instead, the term ``separation probability'' should be used, because although $S_{\text{rev}}(t\vert r_{0})$ is the probability that the molecule is not bound at time $t$, this does not necessarily imply that it was not bound before $t$ and dissociated again, in contrast to the irreversible case \cite{Agmon:1990p10, TPMMS_2012JCP}. Nevertheless, the notion ``survival probability'' is still widely used also in the context of the reversible reaction \cite{kimShin:1999}. 
We adhere to this tradition and shall use both terms interchangeably. 
\subsection{Initially unbound state}
\label{sec-3-1}

Next, we aim to find an expression for the reaction rate $R_{\text{rev}}(t\vert r_{0})$ in terms of $S^{<}_{\text{rev}}(t\vert r_{0})$. To this end, we integrate Eq.~(\ref{rev_FKE_1}) over all space, multiply the equation by $2\pi$ and thus get
  \begin{equation}\label{eq_S_rev}
    \frac{\partial S_{\text{rev}}(t \vert r_{0})}{\partial t}  = -\kappa_r S^{<}_{\text{rev}}(t \vert r_{0})  + \kappa_{d}[1 - S_{\text{rev}}(t\vert r_{0})].
  \end{equation}
The Laplace transform of this equation may be rewritten by virtue of the relation Eq.~(\ref{eq_def_R_laplace}), which remains valid in the reversible case, provided that one considers the initially unbound molecule. Thus, the reaction rate in the Laplace domain comes out to be 
  \begin{equation}\label{R_and_area_rev}
    \tilde{R}_{\text{rev}}(s \vert r_{0})  = \frac{\kappa_{r}s}{s+\kappa_{d}}\tilde{S}^{<}_{\text{rev}}(s \vert r_{0}).
  \end{equation}
We observe that Eq.~(\ref{R_and_area_rev}) assumes the same structure as its irreversible counterpart (Eq.~(\ref{R_and_area})). In fact, the sole difference lies in the form of the recombination rate $\kappa_{r}s/(s+\kappa_{d})$ that becomes dependent on $s$ in the reversible case. In the limit $\kappa_{d} \rightarrow 0$, the recombination rate reduces to $\kappa_{r}$, as it should. This is quite reminiscient of the situation in the CR model, cf. Ref.~\cite[Eq.(3.3b)]{Agmon:1990p10}.

We now turn to the equation of motion for the survival probability that may be derived upon applying the same procedure we already used in Sec.~\ref{sec-2}. However, we have to be aware of the fact that in general
  \begin{equation}
    q(r, t \vert r_{0}) \neq q(r_{0}, t \vert r).  
  \end{equation}
This issue can easily be dealt with by solving Eq.~(\ref{rev_FKE_2}) and expressing $q_{\text{rev}}(r, t\vert r_{0})$ in terms of $p_{\text{rev}}(r, t\vert r_{0})$ 
  \begin{equation}
    q_{\text{rev}}(r, t \vert r_{0}) = \kappa_{r}\Theta(a-r)\int^{t}_{0}e^{-\kappa_{d}(t-t')}p_{\text{rev}}(r, t' \vert r_{0}) dt'.
  \end{equation}
Then, employing the detailed balance condition (Eq.~(\ref{detailed_balance})) and interchanging $r\leftrightarrow r_{0}$, we find
    \begin{eqnarray}\label{backward_GF_rev}
  \frac{\partial p_{\text{rev}}(r, t\vert r_{0})}{\partial t} && = \mathcal{L}_{r_{0}}p_{\text{rev}}(r, t\vert r_{0}) \quad\quad\quad \nonumber\\
&& - \kappa_{r}\Theta(a - r_{0})\bigg[p(r, t\vert r_{0}) - \kappa_{d}\int^{t}_{0}e^{-\kappa_{d}(t-t')}p(r, t'\vert r_{0}) dt'\bigg].\quad\quad\quad
    \end{eqnarray}
We integrate over all space to arrive at
  \begin{eqnarray}\label{eq_motion_S_rev}
    \frac{\partial S_{\text{rev}}(t \vert r_{0})}{\partial t} &&  = \mathcal{L}_{r_{0}}S_{\text{rev}}(t \vert r_{0}) \quad\quad\quad\nonumber\\
&& - \kappa_{r}\Theta(a - r_{0})\bigg[S_{\text{rev}}(t \vert r_{0}) - \kappa_{d}\int^{t}_{0}e^{-\kappa_{d}(t-t')} S_{\text{rev}}(t'\vert r_{0}) dt'\bigg].\quad\quad\quad
  \end{eqnarray}
We point out that, as in the irreversible case, one can demonstrate by virtue of Eq.~(\ref{backward_GF_rev}) that both $S^{<}_{\text{rev}}(t\vert r_{0})$ and $S^{>}_{\text{rev}}(t\vert r_{0})$ obey the same equation of motion as $S_{\text{rev}}(t\vert r_{0})$ (Eq.~(\ref{eq_motion_S_rev})) and that again they are subject to different IC, cf. Eqs.~(\ref{IC_S_<}),~(\ref{IC_S_>}).
Next, the Laplace transform of Eq.~(\ref{eq_motion_S_rev}) yields \cite{Khokhlova:2012} 
  \begin{equation}\label{laplace_eq_motion_S_rev}
    s\tilde{S}_{\text{rev}}(s \vert r_{0}) - 1 = \mathcal{L}_{r_{0}}\tilde{S}_{\text{rev}}(s \vert r_{0}) - \frac{\kappa_{r}s}{s+\kappa_{d}}\Theta(a - r_{0})\tilde{S}_{\text{rev}}(s \vert r_{0}). 
  \end{equation}
We notice that Eq.~(\ref{laplace_eq_motion_S_rev}) enjoys exactly the same form as its irreversible counterpart, provided one makes the by now obligatory substitution $\kappa_{r} \rightarrow \kappa_{r}s/(s+\kappa_{d})$.
Eq.~(\ref{laplace_eq_motion_S_rev}) may be rewritten as 
\begin{equation}\label{laplace_operator_S_rev}
    \mathcal{L}_{r_{0}} \sS =
    \begin{cases}
      Dw^{2} \sS - 1 & \text{if} \quad r_{0} < a \\
      Dv^{2} \sS - 1 & \text{if} \quad r_{0} > a,
    \end{cases}
\end{equation}
where $w$ is now defined by
  \begin{eqnarray}
    w &=& v\sqrt{\frac{s+\kappa_{r}+\kappa_{d}}{s+\kappa_{d}}}. 
  \end{eqnarray}
Obviously, $w$ reduces to the corresponding irreversible expressions Eq.~(\ref{eq_def_w_irr}) for $\kappa_{d} = 0$. By appeal of Eq.~(\ref{eq_def_R_laplace}), the relation for the reaction rate comes out to be 
  \begin{equation}\label{laplace_operator_R_rev}
      \mathcal{L}_{r_{0}} \tilde{R}_{\text{rev}}(s \vert r_{0}) =
      \begin{cases}
        w^{2} \tilde{R}_{\text{rev}}(s \vert r_{0}) - \frac{\kappa_{r}s}{s+\kappa_{d}} & \text{if} \quad r_{0} < a \\
        v^{2} \tilde{R}_{\text{rev}}(s \vert r_{0})  & \text{if} \quad r_{0} > a,
      \end{cases}
  \end{equation}
Note that Eq.~(\ref{laplace_operator_R_rev}) could alternatively have been obtained by
Eq.~(\ref{laplace_operator_R}) via the standard replacement $\kappa_{r} \rightarrow \kappa_{r}s/(s+\kappa_{d})$.
\subsection{Reversible time-dependent rate coefficient}
\label{sec-3-2}

To give meaning to the notion of the reversible time-dependent rate coefficient within the framework of the AR model, we can proceed in analogy to the case of CR theories and to the irreversible case within the AR framework. 
In fact, Eqs. (\ref{kI_by_R}), (\ref{kI_by_dt_S}) and (\ref{kI_by_dr_S}) are unique in the sense that they remain valid in exactly the same form for the irreversible as well reversible reaction within both the CR and AR framework, the only necessary replacements consist of straighforward switching $\kI\rightarrow k_{\text{rev}}(t),\, S(t\vert r_{0}) \rightarrow  S_{\text{rev}}(t\vert r_{0})$. 
In this context, we would like to point out that, for the reversible reaction, Eq.~(\ref{kI_by_dr_S}) results from integrating Eq.~(\ref{eq_motion_S_rev}), instead of Eq.~(\ref{eq_motion_S}), over $r_{0}$ from $a$ to $\infty$.

We remind ourselves, that it is also possible to relate $k_{\text{rev}}(t)$ with the survival probability at contact \cite[Eq.(3.5)]{Agmon:1990p10}
  \begin{equation}\label{kR_cr_Sa}
    \tilde{k}_{\text{rev}}(s) = \frac{s\kappa_{a}}{s+\kappa_{d}} \tilde{S}_{\text{rev}}(s\vert a)
  \end{equation}
Analogously, we find 
  \begin{eqnarray}
    \tilde{k}_{\text{rev}}(s) &=& \frac{s\kappa_{r}}{s+\kappa_{d}} \tilde{S}^{<}(s \vert \mathbf{B}^{C}_{2}) \\
        &=& \frac{s\kappa_{r}}{s+\kappa_{d}} \tilde{S}^{>}(s \vert \mathbf{B}_{2}), \label{k_rev_S}
  \end{eqnarray}
where we used Eq.~(\ref{eq_S_rev}).
As already pointed out in Sec.~\ref{sec-2}, the quantities $S^{<}(s \vert \mathbf{B}^{\mathbf{C}}_{2}), S^{>}(s \vert \mathbf{B}_{2})$ do not technically represent probabilities. However, it is again possible to relate these quantities to the probability
  \begin{equation}
    S^{>}(t \vert \, \text{uni}) =   \frac{1}{\pi a^{2}} S^{>}(t \vert \mathbf{B}_{2}),
  \end{equation}
that the molecule is found unbound with $r>a$ at time $t$, given that initially it was uniformly
distributed inside the reaction area.
Hence, we may write
  \begin{eqnarray}
    \tilde{k}_{\text{rev}}(s) = \frac{s\pi a^{2}\kappa_{r}}{s+\kappa_{d}} \tilde{S}^{>}(s \vert \, \text{uni}). \label{k_rev_S_uni}
  \end{eqnarray} 
We will later employ this relation to give a simple proof that $\int^{\infty}_{0}k(t)dt = K_{\text{eq}}$.

Finally, we would like to point out the relation between $k_{\text{rev}}(t)$ and the time-dependent rate coefficient associated with absorbing BC. We invoke the same line of reasoning given already in the irreversible context, cf. Sec.~\ref{sec-3} and Ref.~\cite{Agmon:1990p10}. Consequently, one has the convolution relation 
  \begin{equation}\label{conv_R_rev}
    R_{\text{rev}}(t \vert r_{0}) = \int^{t}_{0} R_{\text{rev}}(t-t' \vert a)R_{\text{abs}}(t' \vert r_{0}) dt',  
  \end{equation}
that yields
  \begin{equation}
    \frac{\partial \tilde{S}_{\text{rev}}(s \vert r_{0})}{\partial r_{0}}\vert_{r_{0}=a} = \tilde{R}_{\text{rev}}(s \vert a)\frac{\partial\tilde{S}_{\text{abs}}(s \vert r_{0})}{\partial r}\vert_{r_{0}=a},  
  \end{equation}
and hence 
  \begin{equation}
    \tilde{k}_{\text{rev}}(s) = \tilde{R}_{\text{rev}}(s\vert a)\tilde{k}_{\text{abs}}(s). 
  \end{equation}
\subsection{Initially bound state}
\label{sec-3-3}

We now focus on the initially bound state. One can show that the GF solutions corresponding to the
initially bound and unbound state are related by \cite{Khokhlova:2012, Prustel_Area:2014}
\begin{equation}
\tilde{p}(r,s \vert r_{0}, \ast) = \frac{\kappa_{d}}{s+\kappa_{d}}\tilde{p}(r,s \vert r_{0}).
\end{equation}
It follows immediately that
  \begin{eqnarray}
    \tilde{S}_{\text{rev}}(s \vert r_{0}, \ast) = \frac{\kappa_{d}}{s+\kappa_{d}}\tilde{S}_{\text{rev}}(s \vert r_{0}), \label{eq_conv_S_ast_S_laplace}\\
    \tilde{S}^{>}_{\text{rev}}(s \vert r_{0}, \ast) = \frac{\kappa_{d}}{s+\kappa_{d}}\tilde{S}^{>}_{\text{rev}}(s \vert r_{0}). \label{eq_conv_S>_ast_S>_laplace}
  \end{eqnarray}
The relation between the separation probabilities of the two IC (bound and unbound state) is well-known from CR models \cite[Eq.~(3.15)]{Agmon:1990p10} and leads to a convolution relation in the time domain
  \begin{equation}
    S_{\text{rev}}(t \vert r_{0}, \ast) = \kappa_{d}\int^{t}_{0}e^{-\kappa_{d}(t-t')}S_{\text{rev}}(t' \vert r_{0})dt'.
  \end{equation}
We now express $\tilde{S}^{>}_{\text{rev}}(s \vert r_{0})$ by $\tilde{S}^{>}_{\text{rev}}(s \vert r_{0}, \ast)$ via Eq.~(\ref{eq_conv_S>_ast_S>_laplace}) and insert the result in Eq.~(\ref{k_rev_S_uni}) to obtain
  \begin{eqnarray}
    \tilde{k}_{\text{rev}}(s) = \frac{s\pi a^{2}\kappa_{r}}{\kappa_{d}} \tilde{S}^{>}(s \vert \,\text{uni}, \ast) \label{k_rev_S_ast}.
  \end{eqnarray} 
Because $S^{>}(t=0\vert \,\text{uni}, \ast) = 0$, this relation yields in the time domain
  \begin{eqnarray}
    k_{\text{rev}}(t) = \frac{\pi a^{2}\kappa_{r}}{\kappa_{d}} \frac{\partial}{\partial t}S^{>}(t \vert \,\text{uni}, \ast) \label{k_rev_S_ast_time_domain}.
  \end{eqnarray} 
Hence, we can easily conclude that
  \begin{equation}
    \int^{\infty}_{0}k(t) dt = \frac{\pi a^{2}\kappa_{r}}{\kappa_{d}} = K_{\text{eq}},
  \end{equation}
cf. Ref.~\cite{Agmon:1990p10, Prustel_Area:2014}.
\subsection{Asymptotic expressions}
\label{sec-3-4}

Classically, the average lifetime of the bound state is defined by \cite{Agmon:1990p10} 
  \begin{equation}
    \tau_{\text{off}} = \int^{\infty}_{0}[1 - S_{\text{rev}}(t \vert \ast)]dt = \int^{\infty}_{0}Q_{\text{rev}}(t \vert \ast)dt.
  \end{equation}
The macroscopic off-rate is related to average lifetime of the bound state via
  \begin{equation}
    k_{\text{off}} = \frac{1}{\tau_{\text{off}}}.
  \end{equation}
It is well known that in 2D the average lifetime diverges, if one considers the infinite plane, cf. for instance \cite{TPMMS_SP:2011, TPMMS_Note:2012} and references given therein.
To address this issue, one may first consider the average lifetime  up to a certain time $t$
  \begin{equation}\label{eq_tau_off_finite_CR}
    \tau_{\text{off}}(t) = \int^{t}_{0}[1 - S_{\text{rev}}(t' \vert \ast)]dt' = \int^{t}_{0}Q_{\text{rev}}(t' \vert \ast)dt',
  \end{equation}
which is finite for all $t < \infty$. Then, one can apply a large time expansion to $\tau_{\text{off}}(t)$ to analyze the type of the singularity and to separate finite and singular contributions. 

Within the context of AR models, we expect the average lifetime of the bound state to diverge also. Therefore, we adopt a similar strategy and consider the average lifetime  up to a certain time $t<\infty$. However, we cannot directly employ Eq.~(\ref{eq_tau_off_finite_CR}), because we have to take into account that there are infinitely many bound states in AR models, labeled by $r_{0} < a$. Hence, it is natural to define
  \begin{eqnarray}
    \tau_{\text{off}}(t\vert r_{0},\ast) &=& \int^{t}_{0}[1 - S_{\text{rev}}(t' \vert r_{0}, \ast)]dt' = \int^{t}_{0}Q_{\text{rev}}(t' \vert r_{0}, \ast)dt', \quad\label{tau_off_r0}\\
    \tau_{\text{off}}(t) &=& \tau_{\text{off}}(t\vert \text{uni},\ast) = \frac{2}{a^{2}} \int^{a}_{0}\tau_{\text{off}}(t\vert r_{0}) r_{0} dr_{0}. \label{tau_off}
  \end{eqnarray}
Because we are interested in the large time expansion of Eqs.~(\ref{tau_off_r0}),~(\ref{tau_off}), we switch to the Laplace domain and use the explicit expressions derived in Ref.~\cite{Prustel_Area:2014}.
Thus, we obtain
  \begin{eqnarray}
    \tilde{\tau}_{\text{off}}(s \vert r_{0}) & = & \frac{1}{s}\tilde{Q}_{\text{rev}}(s \vert r_{0}, \ast) \\ \label{laplace_tau_Q}
                                        & = & \frac{1}{s(s+\kappa_{d})} + \frac{2\pi \kappa_{r}\kappa_{d}}{s(s+\kappa_{d})^{2}}\int^{a}_{0} \tilde{p}_{\text{rev}}(r, s \vert r_{0}) r dr,
  \end{eqnarray}
and 
  \begin{equation}
    \int^{a}_{0}\tilde{p}_{\text{rev}}(r, s \vert r_{0}) r dr = \frac{1}{2 \pi D w^{2}}\bigg[1 - v\frac{K_{1}(va)I_{0}(wr_{0})}{\mathcal{N}}\bigg].
  \end{equation}
A small $s$ expansion leads us to
  \begin{equation}\label{laplace_tau_expansion}
\frac{1}{2 \pi D w^{2}}\bigg[1 - v\frac{K_{1}(va)I_{0}(wr_{0})}{\mathcal{N}}\bigg] \underset{s\rightarrow 0}{\rightarrow}   \frac{1}{2\pi D}\bigg[\frac{1}{4}(a^{2}-r^{2}_{0}) - \frac{1}{4}a^{2}\ln\bigg(\frac{1}{4}e^{2\gamma}\frac{a^{2}}{D} s \bigg) \bigg]
  \end{equation}
Now we can make use of \cite[Eq.(29.3.98)]{abramowitz1964handbook} 
  \begin{equation}
    \mathbf{L}^{-1} \bigg[\frac{1}{s}\ln s \bigg] = - \gamma - \ln t,
  \end{equation}
where $\gamma$ again refers to the Euler-Mascheroni constant \cite{abramowitz1964handbook}, 
to arrive at the corresponding large time expansion
  \begin{eqnarray}
    \tau_{\text{off}}(t \vert r_{0})& \underset{t\rightarrow \infty}{\rightarrow} & \frac{1}{\kappa_{d}} + \frac{1}{D}\frac{\kappa_{r}}{\kappa_{d}}\bigg[ \frac{1}{4}(a^{2}-r^{2}_{0}) + \frac{1}{4}a^{2}\ln\bigg(4e^{-\gamma}\frac{D}{a^{2}} t \bigg) \bigg],\quad \quad\\
    \tau_{\text{off}}(t) &\underset{t\rightarrow \infty}{\rightarrow}& \frac{1}{\kappa_{d}} + \frac{1}{D}\frac{\kappa_{r}}{\kappa_{d}}\bigg[ \frac{1}{8}a^{2} + \frac{1}{4}a^{2}\ln\bigg(4e^{-\gamma}\frac{D}{a^{2}} t \bigg) \bigg]. \label{eq_large_time_tau_off}
  \end{eqnarray}
We find that the obtained expression is similar to the one obtained predicted by CR theories \cite{TPMMS_Note:2012}. However, one important difference is that the average lifetime depends much stronger, quadratically, on the encounter radius, in contrast to the weak logarithmic dependence observed the classical theory.

Note that although both $1/\kI$ and $\tau_{\text{off}}(t)$ diverge for $t\rightarrow \infty$, one has
  \begin{equation}
  \lim_{t\rightarrow\infty} \frac{\kI}{k_{\text{off}}(t)} = \frac{\pi a^{2 } \kappa_r}{\kappa_{d}} = K_{\text{eq}},
  \end{equation}
because the logarithmic divergence gets cancelled, as one can infer from Eqs.~(\ref{eq_large_time_k_irr}) and (\ref{eq_large_time_tau_off}).

Finally, we turn to the large time approximation of the binding probability of the initially bound state. We employ Eqs.~(\ref{laplace_tau_Q}) and (\ref{laplace_tau_expansion}), furthermore we take into account \cite[Ch. 13.6, Eq.(8)]{carslaw1986conduction}
  \begin{equation}
    \mathbf{L}^{-1}\bigg[\ln(bs) \bigg] = -\frac{1}{t},
  \end{equation}
where $b>0$ denotes an arbitrary constant, to derive
  \begin{equation}
    Q_{\text{rev}}(t \vert r_{0}, \ast) \underset{t\rightarrow\infty}{\longrightarrow} \frac{\kappa_{r}}{\kappa_{d}}\frac{a^{2}}{4 D t}.
  \end{equation}
\section*{Acknowledgments}
This research was supported by the Intramural Research Program of the NIH, National Institute of Allergy and Infectious Diseases. 

\begin{thebibliography}{10}

\bibitem{abramowitz1964handbook}
M.~Abramowitz and I.A. Stegun.
\newblock {\em Handbook of Mathematical Functions with Formulas, Graphs, and
  Mathematical Tables}.
\newblock Dover, New York, 1965.

\bibitem{Agmon:1984}
N.~Agmon.
\newblock {\em J. Chem. Phys.}, 81:2811, 1984.

\bibitem{Agmon:1990p10}
N.~Agmon and A.~Szabo.
\newblock {\em J. Chem. Phys.}, 92:5270, 1990.

\bibitem{Barzykin:1993}
A.V. Barzykin and M.~Tachiya.
\newblock {\em J. Chem. Phys.}, 99:9591, 1993.

\bibitem{carslaw1986conduction}
H.S. Carslaw and J.C. Jaeger.
\newblock {\em Conduction of Heat in Solids}.
\newblock Clarendon Press, New York, 1986.

\bibitem{collins1949diffusion}
F.C. Collins and G.E. Kimball.
\newblock {\em J. Colloid Sci.}, 4:425, 1949.

\bibitem{Goodrich:1954}
F.C. Goodrich.
\newblock {\em J. Chem. Phys.}, 22:588, 1954.

\bibitem{Goesele:1984}
U.M. G\"osele.
\newblock {\em Prog. React. Kinet.}, 13:63, 1984.

\bibitem{Khokhlova:2012BullKorCS}
S.S. Khokhlova and N.~Agmon.
\newblock {\em Bull. Korean Chem. Soc.}, 33:1020, 2012.

\bibitem{Khokhlova:2012}
S.S. Khokhlova and N.~Agmon.
\newblock {\em J. Chem. Phys.}, 137:184103, 2012.

\bibitem{kimShin:1999}
H.~Kim and K.J. Shin.
\newblock {\em Phys. Rev. Lett.}, 82:1578, 1999.

\bibitem{Erban:2011}
J.~Lipkov\'{a}, K.C. Zygalakis, S.J Chapman, and R.~Erban.
\newblock {\em SIAM J. Appl. Math.}, 71:714, 2011.

\bibitem{TPMMS_SP:2011}
T.~Pr{\"u}stel and M.~Meier-Schellersheim.
\newblock arXiv:1112.4010v1 [math-ph], 2011.

\bibitem{TPMMS_2012JCP}
T.~Pr\"ustel and M.~Meier-Schellersheim.
\newblock {\em J. Chem. Phys.}, 137:054104, 2012.

\bibitem{TPMMS_Note:2012}
T.~Pr{\"u}stel and M.~Meier-Schellersheim.
\newblock arXiv:1210.1265v1 [math-ph], 2012.

\bibitem{Prustel_Area:2014}
T.~Pr\"{u}stel and M.~Meier-Schellersheim.
\newblock {\em J. Chem. Phys.}, 140:114106, 2014.

\bibitem{Rice:1985}
S.~A. Rice.
\newblock {\em Diffusion Limited Reactions}.
\newblock Elsevier, New York, 1985.

\bibitem{smoluchowski:1917}
M.~von Smoluchowski.
\newblock {\em Z. Phys. Chem.}, 92:129, 1917.

\bibitem{Wilemski:1973}
G.~Wilemski and M.~Fixman.
\newblock {\em J. Chem. Phys.}, 58:4009, 1973.

\end{thebibliography}

\end{document}